\documentclass[review]{elsarticle}

\usepackage{lineno,hyperref}
\usepackage{graphicx,psfrag}
\usepackage{epstopdf}
\usepackage{amsmath}
\usepackage{bm}
\usepackage{amssymb}
\usepackage{subfigure}
\usepackage{dcolumn}
\usepackage{color}
\usepackage{mathrsfs}
\modulolinenumbers[5]

\journal{Journal of Sound and Vibration}
\begin{document}

\title{Traveling and standing thermoacoustic waves in solid media}

\author[mymainaddress,mysecondaryaddress]{Haitian Hao}
\ead{haoh@purdue.edu}

\author[mymainaddress]{Carlo Scalo}
\ead{scalo@purdue.edu}

\author[mymainaddress,mysecondaryaddress]{Fabio Semperlotti\corref{mycorrespondingauthor}}
\cortext[mycorrespondingauthor]{Corresponding author}
\ead{To whom correspondence should be addressed: fsemperl@purdue.edu}

\address[mymainaddress]{School of Mechanical Engineering, Purdue University, West Lafayette, Indiana 47907, USA}
\address[mysecondaryaddress]{Ray W. Herrick Laboratories, 177 South Russell Street, West Lafayette, Indiana 47907, USA}

\begin{abstract}
The most attractive application of fluid-based thermoacoustic (TA) energy conversion involves traveling wave devices due to their low onset temperature ratios and high growth rates. Recently, theoretical and numerical studies have shown that thermoacoustic effects can exist also in solids. However, these initial studies only focus on standing waves. This paper presents a numerical study investigating the existence of self-sustained thermoelastic oscillations associated with traveling wave modes in a looped solid rod under the effect of a localized thermal gradient. Configurations having different ratios of the rod radius $R$ to the thermal penetration depth $\delta_k$ were explored and the traveling wave component (TWC) was found to become dominant as $R$ approaches $\delta_k$. The growth-rate-to-frequency ratio of the traveling TA wave is found to be significantly larger than that of the standing wave counterpart for the same wavelength. The perturbation energy budgets are analytically formulated and closed, shedding light onto the energy conversion processes of solid-state thermoacoustic (SSTA) engines and highlighting differences with fluids. Efficiency is also quantified based on the thermoacoustic production and dissipation rates evaluated from the energy budgets.
\end{abstract}


\maketitle



\section{\label{sec:level1}Introduction}
Thermoacoustic (TA) instability is a thermodynamic process through which heat is converted into mechanical energy \cite{Rayleigh}. When the working medium is a fluid, this process can be driven by combustion \cite{Poinsot} or, more simply, by wall heat transfer \cite{Rijke}. In both cases, a two-way coupling between fluid motion and fluctuations in heat release rates is established, effectively resulting in a thermodynamic cycle where the fluid parcel produces mechanical (acoustic) work. This inherently cyclic process makes pressure and velocity oscillations grow unbounded in the absence of losses. Recently, Hao \textit{et al.} \cite{Hao} have theoretically demonstrated that this process can also occur with elastic waves in solid media. They provided theoretical and numerical evidence of the existence of thermoacoustic instability in solids by showing unbounded standing wave oscillations in a quarter-wavelength (fixed-free) and a sub-quarter-wavelength (fixed-mass) metal rod. 

The present manuscript provides two key contributions: 1) it extends the concept of solid-state thermoacoustics (SS-TA) to traveling wave configurations, and 2) offers an in-depth analysis of the wave energy budgets of SS-TA. Thermoacoustic instability, in fact, can be exploited to design energy conversion devices called thermoacoustic engines (TAEs) \cite{Swift2}, which are categorized into two types: standing-wave and traveling-wave engines. The difference between them lies in the phase difference between pressure and velocity oscillations. In a standing wave device, the phase difference is approximately (but not exactly equal to) $90^\circ$ at all spatial locations, while in a traveling wave engine it stays well below $90^\circ$ depending on the specific design (\textit{e.g.} between $\pm30^\circ$ in the traveling wave TAE built by Yazaki \cite{Yazaki}). The efficiency is greatly affected by the relative phasing of the oscillations.  Ceperley \cite{Ceperley} was the first to propose that a very efficient pistonless Stirling-like thermodynamic cycle could be achieved with traveling waves propagating through a solid boundary with thermal gradient. Such an engine was experimentally designed by Yazaki \textit{et al} \cite{Yazaki} although at a relatively low efficiency compared to Ceperley's theoretical expectations \cite{Ceperley}. Backhaus and Swift \cite{Backhaus} later designed a new type of traveling-wave TAE based on a compact acoustic network. The addition of a resonator superimposes standing waves on the traveling wave in the engine to decrease the large loss observed in both Ceperley's and Yazaki's designs.

While exhibiting higher thermoacoustic growth rates, traveling-wave TAEs suffer from nonlinear losses such as Gedeon Streaming and other forms of acoustic streaming \cite{Gedeon, Ju, Ravex, Olson, Boluriaan, Scalo}, found to be the main cause of efficiency drop. 

In this study, we prove the existence of traveling thermoacoustic waves in solid media based on the theoretical framework developed previously by the same authors \cite{Hao}. We also show that the growth-rate-to-frequency ratio (shorten as growth ratio hereinafter) of the traveling wave oscillations is considerably larger than that of a standing wave oscillation of the same wavelength. Heat flux, mechanical power, and work source for theoretical solid-state thermoacoustic (SSTA) engines are defined heuristically in light of their definitions in fluids. The acoustic energy budgets are analyzed in detail to interpret the energy conversion process in SSTA engines and to define the efficiencies of SSTA engines. Through the detailed study and comparison of traveling and standing wave thermoacoustics, this paper expands the theory of thermoacoustics of solids and may lead to implementations of new generations of ultra-compact and robust SSTA devices capable of direct thermal-to-mechanical energy conversion.

\section{\label{sec:level1}Problem statement}
\begin{figure}[h]
	\centering
	\includegraphics[scale=0.25]{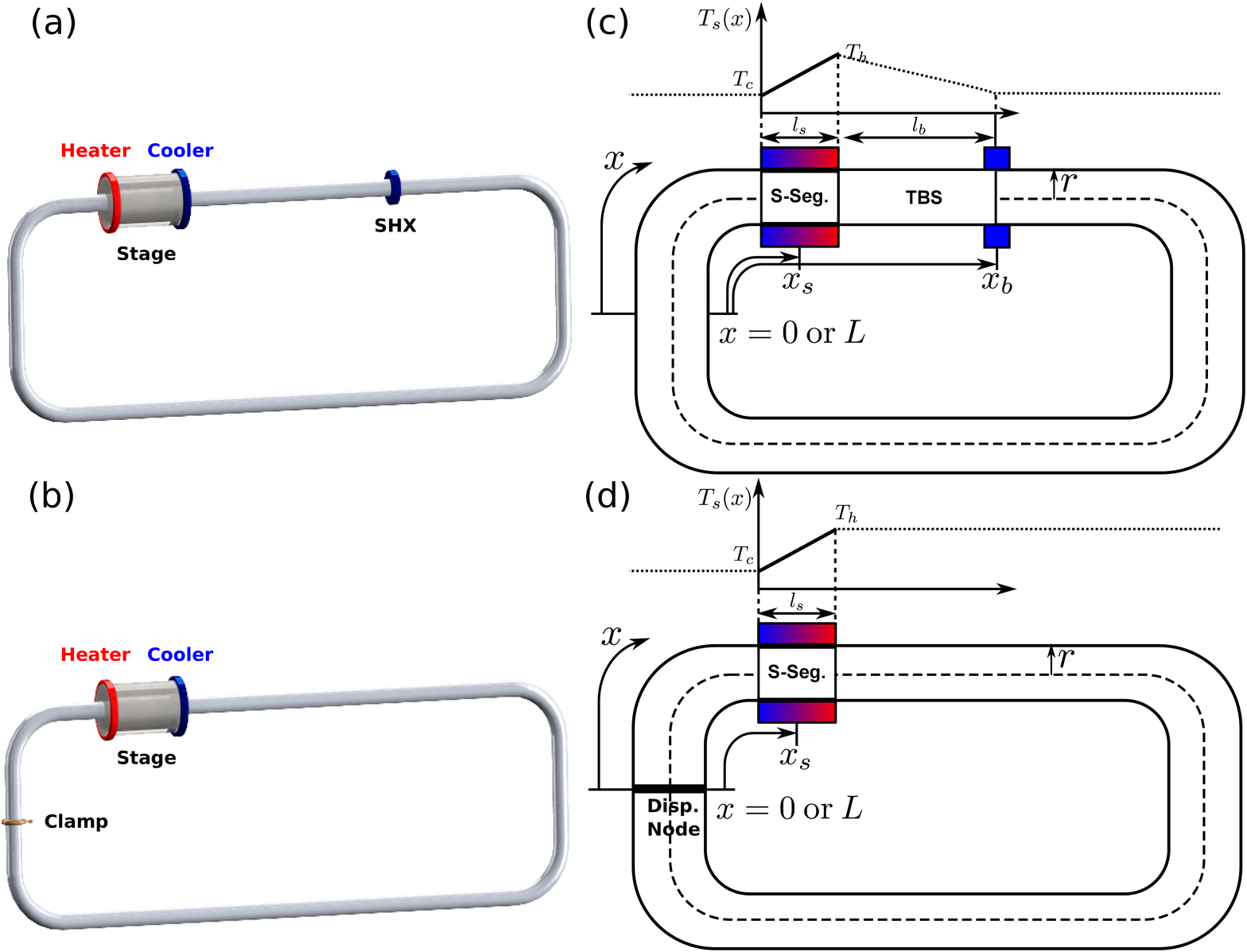}
	\caption{Notional schematics of (a) the looped rod and (b) the resonance rod. A component with a large thermal inertia, \textit{stage}, connected to a heater and a cooler on opposite ends, is mounted on the outer surface of the rod to sustain a linear thermal gradient. In (a), a secondary cold heat exchanger (SHX) is attached to the rod creating the Thermal Buffer Segment (TBS, shown in (c)). In (b), a clamp is used to apply the displacement node (abbreviated as Disp. Node in (d)), which is necessary to suppress the traveling wave mode. (c) and (d) show the temperature profile $T_0(x)$ in the S-seg. (solid line, $T_s(x)$), and in the remaining sections (dashed line), and the characteristic geometric parameters. $T_h$ and $T_c$ are the hot and cold temperatures respectively. The stage is $l_s=0.05L$ long centered about $x=x_s$ (irrelevant for the looped design). The SHX is mounted at $x_b$ ($l_b=0.45L$ away from the stage). The optimal location of the stage's midpoint $x_s$ for the full-wavelength standing wave is $x_s=0.845L$. }
	\label{sketch}
\end{figure}

In this study, we consider two configurations (Fig. \ref{sketch}) in which a ring-shaped slender metal rod with circular cross section is under investigation. Specifically, they are called the looped rod (Fig. \ref{sketch}(a) and (c)) and the resonance rod (Fig. \ref{sketch}(b) and (d)). The rod experiences an externally imposed axial thermal gradient applied via isothermal conditions on its outer surface at a certain location, while the remaining exposed surfaces are adiabatic. The difference between the two configurations lies in the imposition of a displacement/velocity node (Fig. \ref{sketch}(d)), which is used in the resonance rod to suppress the traveling wave mode. Practically, the displacement node could be realized by constraining the rod with a clamp at a proper location (Fig. \ref{sketch}(b)). The coupled thermoacoustic response induced by the external thermal gradient and the initial mechanical excitation is investigated.

The initial mechanical excitation could grow with time as a result of the coupling between the mechanical and thermal response provided a sufficient temperature gradient is imposed on the outer boundary of a solid rod at a proper location. This phenomenon is identified as the thermoacoustic response of solids in \cite{Hao}. 

By analogy with fluid-based traveling wave thermoacoustic engines \cite{Yazaki,Swift}, a stage element is used to impose a thermal gradient on the surface of the looped rod (Fig. \ref{sketch}(a)). The specific location of the stage element in this case is irrelevant due to the periodicity of the system. The segment surrounded by the stage is named S-segment, which experiences a spatial temperature gradient (from $T_c$ to $T_h$) due to the externally enforced temperature distribution. The interface between the stage and the S-segment is ideally assumed to have a high thermal conductivity, which assures the isothermal boundary conditions along with a zero shear stiffness. One can anticipate the compromise between these two seemingly contradictory conditions in an experimental validation. The stage is considered as a thermal reservoir so that the temperature fluctuation on the surface of S-segment is assumed to be zero (isothermal). A Thermal Buffer Segment (TBS) next to the thermal gradient provides a thermal buffer between $T_h$ and room temperature $T_c$. The temperature drop in the TBS is caused by the secondary cold heat exchanger (SHX, Fig. \ref{sketch}(a)) located at $x_b$. A linear temperature profile in the TBS from $T_h$ to $T_c$ is adopted to account for the natural axial thermal conduction along the looped rod. 

\begin{figure}[h]
	\centering
	\includegraphics[scale=0.2]{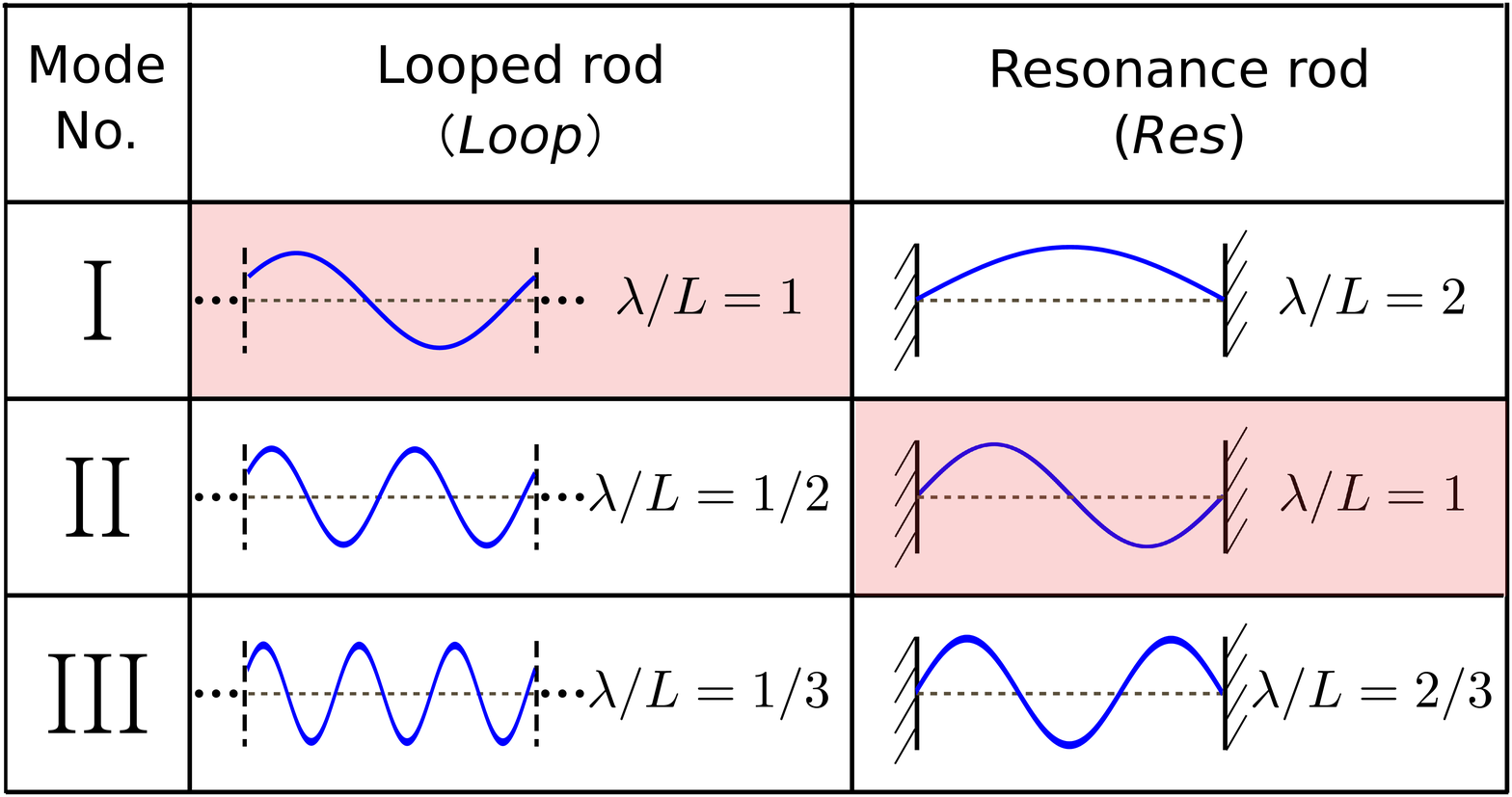}
	\caption{The mode shapes of the looped and the resonance rod and the naming convention for modes. Note that same mode numbers correspond to different wavelengths. Especially, the looped rod starts with a full-wavelength mode as its first mode while a resonance rod starts with a half-wavelength one. To make a comparison based on the same wavelength, $Loop-I$ and $Res-II$ represent our contrast group (the shaded blocks). }
	\label{modes}
\end{figure}

To show the superiority of traveling wave thermoacoustics, a fair comparison was conducted with a resonance rod. The resonance rod, as Fig. \ref{sketch}(d) shows, was constructed by enforcing a displacement/velocity node at an arbitrary position labeled $x=0$. This node is equivalent to a fixed and adiabatic boundary condition. If only plane wave propagation is considered, this resonance rod has no difference with a straight rod with both ends clamped. The TBS is not necessary in the resonance rod since the temperature can be discontinuous at the displacement node. To make a comparison, we calculated the growth ratio of a standing wave mode in the resonance rod with the same wavelength ($\lambda=L$) and frequency ($\approx2830$Hz) as the traveling wave mode in the looped rod without the displacement node. We highlight the essential difference of the mode numbering in Fig. \ref{modes} and propose a naming convention for the modes for brevity. The modes in comparison in this study are $Loop-I$ and $Res-II$ (the shaded blocks).

With the boundary conditions well defined, the governing equations can be solved to show the transient thermoacoustical response of the system. 

\section{\label{sec:level1}Mathematical modeling}

The laws of thermoelasticity are considered to model thermoacoustics in solids in that an elastic wave propagating in a solid medium, whether growing or decaying, is accompanied by a thermal wave. The essential difference with previous studies in thermoelasticity is the presence of heat exchange between the solid medium and its boundary. Hao \textit{et al.} \cite{Hao} discovered that thermoelastic waves can be made thermoacoustically unstable. In the following, we analyze the thermoacoustic response of the setup in Fig. \ref{sketch} adopting the previously developed thermoacoustic linear stability model \cite{Hao} according to Rott's theory \cite{Rott}.

The linearized analysis is performed around mean state $\{u_0, v_0, T_0\}=\{0, 0, T_0\}$, where the subscript $0$ denotes {\it{base state}} because they are zero order terms. The solid is assumed to be homogeneous and isotropic. The first order fluctuating terms (with subscript `1') are assumed to be harmonic in time, namely $()_1=()-()_0=\hat{()}e^{i\Lambda t}$, where $\hat{()}$ refers to the fluctuating variable in the frequency domain, $\Lambda=-i\beta+\omega$, $\omega$ is the angular frequency of the harmonic response, and $\beta$ is the growth rate. The linearized quasi-1D equations are written as
\begin{align}
i\Lambda \hat{u}&=\hat{v}, \label{1dint}\\
i\Lambda \hat{v}&=\frac{E}{\rho}\bigg(\frac{d^2\hat{u}}{dx^2}-\alpha\frac{d\hat{T}}{dx}\bigg), \label{1dmom}\\
i\Lambda \hat{T}&=-\frac{dT_0}{dx}\hat{v}-\gamma_G T_0\frac{d\hat{v}}{dx}+i\omega g_k \hat{T}, \label{1dene}
\end{align}
where $i$ is the imaginary unit, $\hat{u}$, $\hat{v}$ and $\hat{T}$ are the fluctuations of the particle displacement, particle velocity, and temperature averaged over the cross section of the rod, $\gamma_G=\frac{\alpha E}{\rho c_\epsilon}$ is the one-dimensional $Gr\ddot{u}neisen$ constant \cite{Yates}. The dimensionless function $g_k$ is given by
\begin{align}
g_k=
\begin{cases}
\dfrac{1}{1-\frac{1}{2}\xi_{top} \dfrac{J_0(\xi_{top})}{J_1(\xi_{top})}} &x_s-\frac{l_s}{2}<x<x_s+\frac{l_s}{2}\\
0 &\text{elsewhere}
\end{cases},
\end{align}
where $J_n(\cdot)$ are Bessel functions of the first kind, $\xi_{top}=\sqrt{-2i}\frac{R}{\delta_k}$ is the dimensionless complex radius, $R$ is the radius of the looped rod. The thermal penetration depth $\delta_k$ is defined as $\delta_k=\sqrt{\frac{2\kappa}{\omega \rho c_\epsilon}}$. This quantity represents the characteristic thermal penetration depth from the isothermal boundary in the radial direction. The temperature fluctuation caused by the heat exchange between the solid media and SHX is neglected considering the small size of SHX. As a result, $g_k(x_b)=0$. 

An eigenvalue analysis  $(i\Lambda {\bf I} - {\bf A} ){\bf y} = {\bf 0}$ was performed based on the linear quasi-1D model to find the angular frequency $\omega$ and growth rate $\beta$. In the eigenvalue problem, ${\bf I}$, ${\bf A}$ and $\textbf{0}$ are the identity matrix, coefficient matrix, and the null vector respectively, and ${\bf y}=[{\bf\hat{u}};{\bf\hat{v}};{\bf\hat{T}}]$ is the vector of state variables where ${\bf\hat{u}},{\bf\hat{v}}$, and ${\bf\hat{T}}$ are the eigenfunctions of $\hat{u}$, $\hat{v}$, and $\hat{T}$.

\section{\label{sec:level1}Results}
We solved the eigenvalue problem numerically for both cases of a $L=1.8$m long aluminum rod, being the looped or the resonance rod, under a $200$K temperature difference ($T_h=493.15$K and $T_c=293.15$K) with a $0.05L$ long stage to investigate the thermoacoustic response of the system. The material properties of aluminum are chosen as: Young's modulus $E=70$GPa, density $\rho=2700$kg/m$^3$, thermal expansion coefficient $\alpha=23\times 10^{(-6)}$K$^{-1}$, thermal conductivity $\kappa=238$W/(m$\cdot$K) and specific heat at constant strain $c_\epsilon=900$J/(kg$\cdot$K). 

\begin{figure}[h]
	\centering
	\includegraphics[scale=0.6]{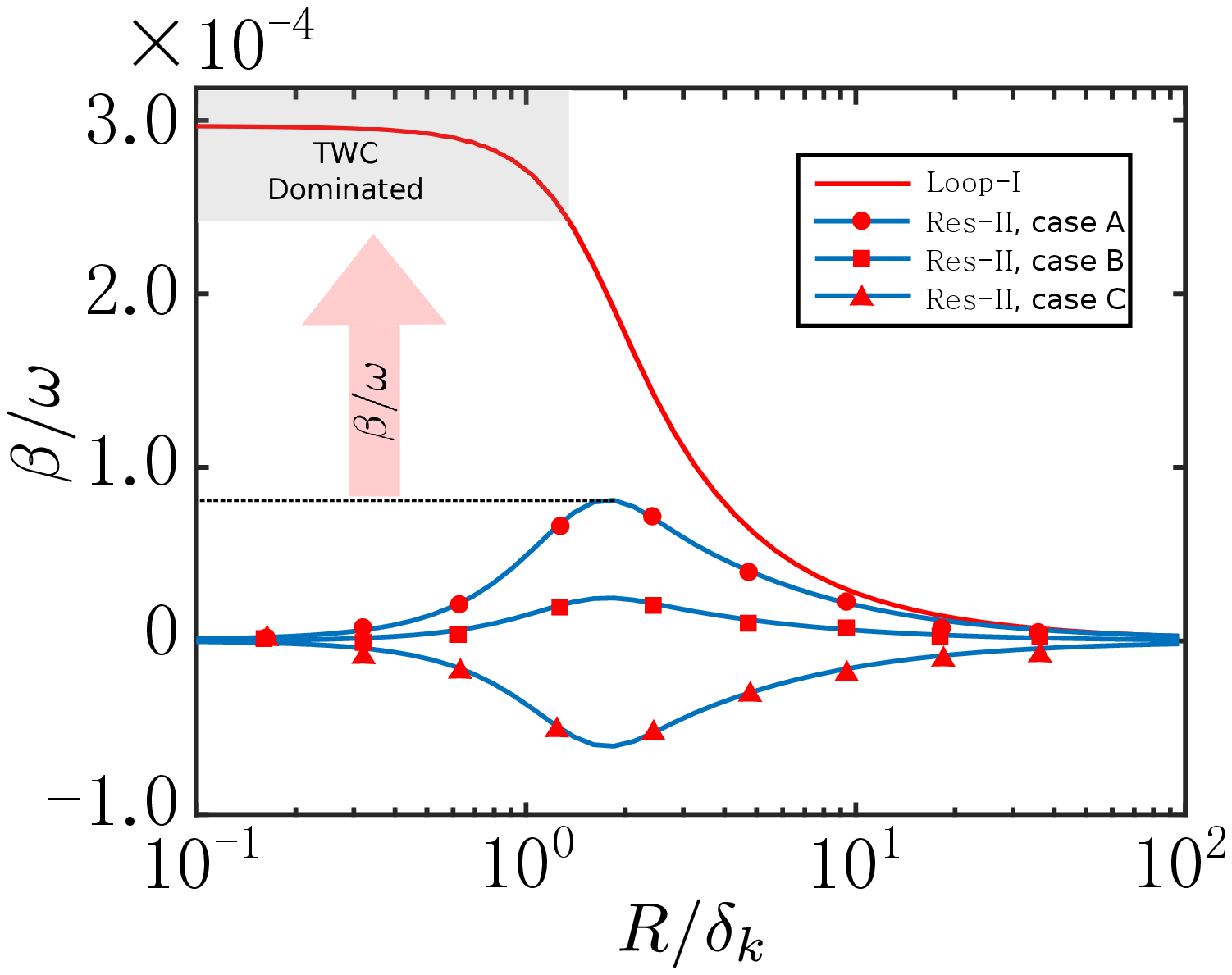}
	\caption{A semilog plot of the growth ratio versus the nondimensional radius for the $Loop-I$ mode in the looped rod and the $Res-II$ mode in the resonance rod. Case A, B, C correspond to $Res-II$ mode with the stage placed at different locations. The growth ratios of these three cases at optimal $R/\delta_k$ are plotted in Fig. \ref{fixed}}
	\label{ratio}
\end{figure}

The first traveling wave mode in the looped rod, with a full wavelength $(\lambda=L)$ is considered, and will be referred to as $Loop-I$, following the naming convention of modes shown in Fig. \ref{modes}. The dimensionless growth ratio $\beta/\omega$ is used as the merit for the SSTA engine's ability to convert heat to mechanical energy. The optimal growth ratio was found by gradually varying the radius $R$ of the looped rod. We used the dimensionless radius $R/\delta_k$ to represent the effect of geometry, where $\delta_k$ was assumed to be constant at the operating frequency $f=\frac{c}{\lambda}\approx\frac{\sqrt{E/\rho}}{L}=2830$Hz. The `$Loop-I$' curve in Fig. \ref{ratio} shows the growth ratio $\beta/\omega$ vs. the dimensionless radius  $R/\delta_k$ of a full-wavelength traveling wave mode. The frequency variation with radius is neglected. Positive growth ratios are found in the absence of losses, which in solids are mainly induced by structural damping. The positive growth ratio suggests that the undamped system is capable of sustaining and amplifying the propagation of a traveling wave.

\begin{figure}[h]
	\centering
	\includegraphics[scale=0.6]{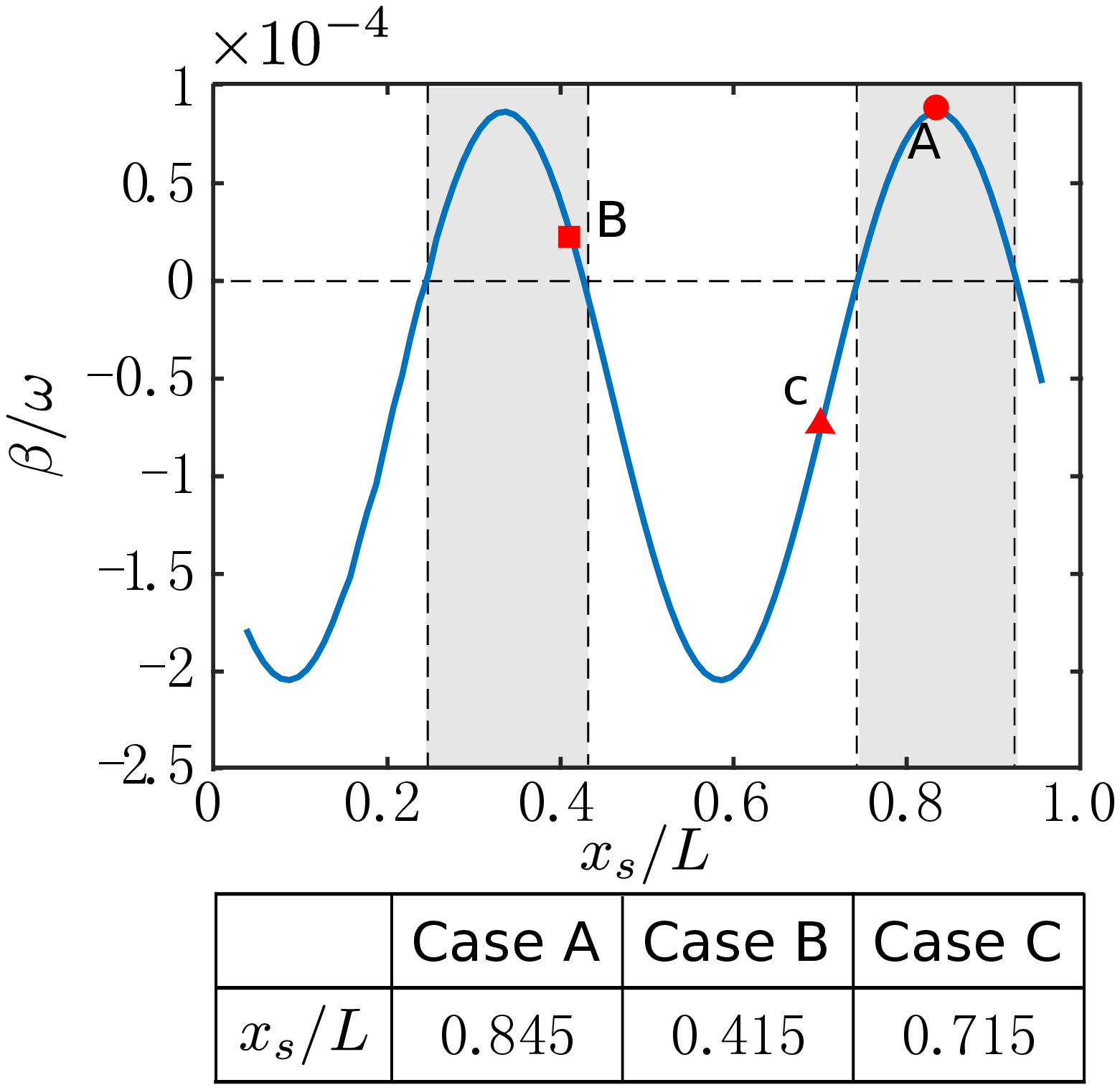}
	\caption{Plot of the growth ratio versus the normalized stage location for the resonance rod $Res-II$ at optimal $R/\delta_k(=1.8)$. Three specific cases are labeled A, B and C corresponding to the stability curves in Fig. \ref{ratio}. Only the location of the stage falling into the shaded region gives a positive growth ratio.}
	\label{fixed}
\end{figure}

On the other hand, for the resonance rod configuration, only standing-wave thermoacoustic waves can exist since the traveling wave mode is suppressed by the displacement node. In this case, the second mode (also $(\lambda=L)$) is considered, and denoted as $Res-II$ (Fig. \ref{modes}) The presence of a displacement node also decreases the rod's degree of symmetry. Thus, the stage location, while being irrelevant in the looped rod configuration, crucially affects the growth ratio in the standing wave resonance rod. An improper placement of the stage on a resonance rod can lead to a negative growth rate, physically attenuating the oscillations. As Fig. \ref{fixed} shows, only a proper location falling into the shaded region leads to a positive growth ratio. Other than the stage location, the radius of the rod is also another important factor, which can affect the growth ratio for the resonance rod configuration. In Fig. \ref{ratio}, we show the $\beta/\omega$ vs. $R/\delta_k$ relations of a resonance rod for different stage locations as well. The maximum thermoacoustic response is obtained for a stage location $x_s=0.845L$ ($Res-II$,case A).

Figure \ref{ratio} shows that as $R\gg\delta_k$, all the curves, whether the looped or the resonance rod, reach zero due to the weakened thermal contact between the solid medium and the stage. However, as $R/\delta_k$ reaches zero (shaded grey region), the stage is very strongly thermally coupled with the elastic wave. As a result, the traveling wave mode dominates. The stability curves also tell that the traveling wave engine has about 4 times higher growth ratio in the limit $R/\delta_k\rightarrow 0$, compared to the standing wave resonance rod ($Res-II$,case A) in which maximal growth ratio is obtained (at $R/\delta_k\approx2$). The noteworthy improvement on growth ratio is essential to the design of more robust solid state thermoacoustics devices.

Hereafter, the modes or results from $Loop-I$ and $Res-II$ will be taken for values of $R$ of $0.1$mm and $0.184$mm, \textit{i.e.} $R/\delta_k$ of $1.0$ and $1.8$ respectively.

\begin{figure}[h]
	\centering
	\includegraphics[scale=0.3]{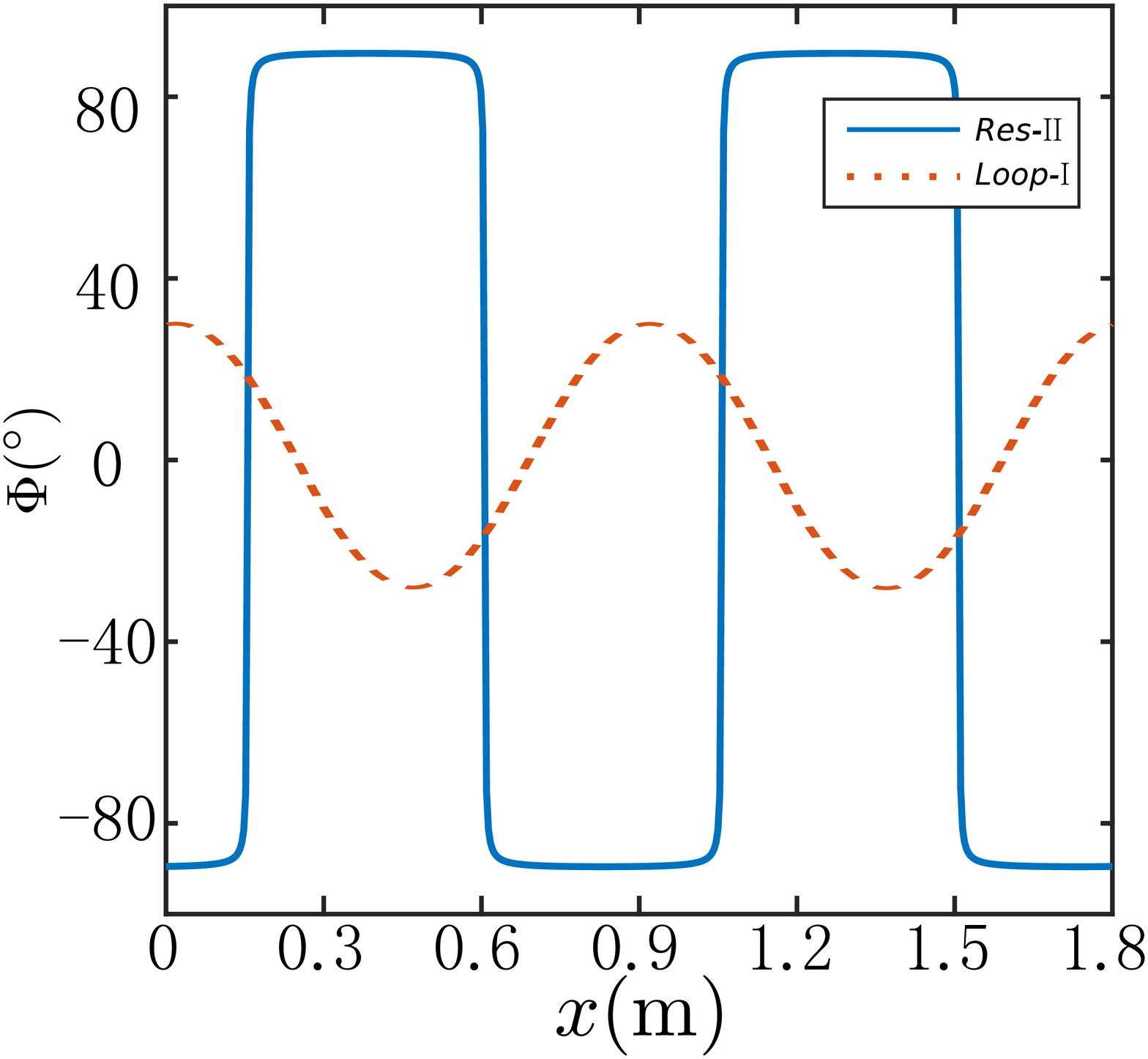}
		\caption{Plot of the phase difference between negative stress $\bar\sigma$ and particle velocity $v$ for a $R=0.184$mm resonance rod `$Res-II$' versus a $R=0.1$mm looped rod `$Loop-I$'.}
	\label{phase}
\end{figure}

In classical thermoacoustics, the phase delay between pressure and cross-sectional averaged velocity is an essential controlling parameter of thermoacoustic energy conversion. In analogy with thermoacoustics in fluids, we use the phase difference $\Phi$ between negative stress $\bar\sigma=-\sigma=\vert\hat{\bar\sigma}\vert \text{Re}[e^{i(\omega t+\phi_{\bar\sigma})}]$ and particle velocity $v=\vert \hat{v}\vert  \text{Re}[e^{i(\omega t+\phi_v)}]$, where $\phi_{\bar\sigma}$ and $\phi_v$ denote the phases of $\bar\sigma$ and $v$ respectively, $\Phi=\phi_v-\phi_{\bar\sigma}$. Note that a negative stress in solids indicates compression which is equivalent to a positive pressure in fluids. The standing wave component (SWC) and traveling wave component (TWC) of velocity are quantified as $v_S=\vert \hat{v}\vert  \text{Re}[e^{i(\omega t+\phi_{\bar\sigma}+\pi/2)}]sin\Phi$ and $v_T= \vert \hat{v}\vert  \text{Re}[e^{i(\omega t+\phi_{\bar\sigma})}]cos\Phi$, which are $90^\circ$ out-of-phase and in-phase with $\bar\sigma$, respectively. In a resonance rod, TWC is not existent. However, the non-zero growth rate $\beta$ will cause a small phase shift, which makes the phase difference $\Phi$ close to but not exactly $90^\circ$. The blue solid line in Fig. \ref{phase} shows the phase difference of a $R=0.184$mm resonance rod ($Res-II$). In the case of a thick looped rod ($R\gg\delta_k$) with a poor degree of thermal contact, the mode shape is much similar to that of a resonance rod because SWC is still dominant and the phase difference is close to $90^\circ$. Supplementary Movie 1 shows that the displacement nodes may exist intrinsically in the system without clamped points. However, when the looped rod is sufficiently thin ($R\sim\delta_k$) the traveling wave component plays a dominant role. Thus, the phase delay decreases to $30^\circ$ at most. The orange dashed line in Fig. \ref{phase} shows the phase difference of a $R=0.1$mm looped rod ($Loop-II$). The time history of the displacement along the looped rod in Supplementary Movie 2 shows that, as $R\leq\delta_k$ (small phase difference), the wave mode is dominated by TWC.

\section{\label{sec:level1} Energy conversions in solid-state thermoacoustic engines}
In this section, we explore the energy conversion process in the resonance and the looped rods. The resonance rod, `$Res$', has a length of $1.8$m, radius of $R=0.184$mm and the stage location $x_s=0.805L$. The looped rod, `$Loop$', has the same total length, but the radius $R=0.1$mm is selected to allow the TWC to dominate. The location of the stage in looped rods does not influence the thermoacoustic response, thus only for illustrative purposes, it is located at $x_s=0.205L$ so that the TBS does not cross the point where periodicity is applied.

First, we adopt heuristic definitions of heat flux and mechanical power (work flux), analogous to the well-defined heat flux and acoustic power in fluids (Section 5.1). The energy budgets are then rigorously derived (Section 5.2), naturally yielding the consistent expressions of the second order energy norm, work flux, energy redistribution term, and the thermoacoustic production and dissipation. The efficiency, the ratio of the net gain (which eventually converts into energy growth) to the total heat absorbed by the medium, is defined based on the acoustic energy budgets and it is found that the first mode of the traveling wave engine  (`$Loop-I$') is more efficient than the second standing wave mode (`$Res-II$'). 

\subsection{\label{sec:level2}Heat flux, Mechanical Energy and Work Source}

A cycle-averaged heat flux in the axial direction is generated in the S-segment due to its heat exchange with the stage. Neglecting the axial thermal conductivity, the transport of entropy fluctuations due to the fluctuating velocity $v_1$ (subscript 1 for a first order fluctuating term in time) is the only way heat can be transported along the axial direction \cite{Swift2}, and it is expressed in the time domain as
\begin{align}
\dot{q_2}=T_0\rho_0(s_1 v_1) \text{ [W/m$^2$]},
\label{hf}
\end{align}
The subscript 2 in the heat flux per unit area $\dot{q_2}$ denotes a second order quantity. Entropy fluctuations in solids are related to temperature and strain rate fluctuations via the following relation from thermoelasticity theory \cite{Biot}:
\begin{align}
s_1=\frac{c_\epsilon}{T_0}T_1+\alpha E\varepsilon_1.
\label{entropy}
\end{align}
Using Eq. (\ref{entropy}) into Eq. (\ref{hf}) , $\dot{q_2}$ can be expressed in terms of $T_1$, $v_1$ and $\varepsilon_1$. The counterparts of these three quantities in frequency domain $\hat{T}$, $\hat{v}$ and $\hat{\varepsilon}$ can be extracted from the eigenfunctions of the eigenvalue problem (Eqs. (\ref{1dint}),(\ref{1dmom}) and (\ref{1dene}). Under the assumption: $\beta/\omega\ll1$, the second order cycle-averaged products $\langle a_1 b_1 \rangle$ can be evaluated as $\langle a_1 b_1 \rangle=\frac{1}{2}\text{Re}[\hat{a}\hat{b}^*]e^{2\beta t}$ (\textit{e.g.} $\langle s_1 v_1\rangle=\frac{1}{2}\text{Re}[\hat{s}\hat{v}^*]e^{2\beta t}$), where $a$ and $b$ are dummy harmonic variables following the $e^{i\Lambda t}$ convention introduced in Section 3, and the superscript $*$ denotes the complex conjugate. We obtain $\langle \dot{q_2}\rangle=\tilde{Q}e^{2\beta t}$, where
\begin{align}
\tilde{Q}=\frac{1}{2}\rho_0 c_\epsilon \text{Re}[\hat{T}\hat{v}^*]+\frac{1}{2}T_0\alpha E \text{Re}[\hat{\varepsilon}\hat{v}^*]\qquad\text{ [W/m$^2$]}.
\end{align}
The total heat flux through the cross section of the rod is 
\begin{align}
\dot{Q}=\int_A \langle \dot{q_2}\rangle dS=A\langle \dot{q_2}\rangle \qquad\text{ [W]}.
\end{align}
The second equality holds because the eigenfunctions are all cross-section-averaged quantities. We note that $\dot{Q}$ is a function of the axial position $x$.

The instantaneous mechanical power carried by the wave is defined as
\begin{align}
I_2=(-\sigma_1)v_1=\bar{\sigma}_1 v_1. \qquad\text{ [W/m$^2$]} \label{MP}
\end{align}
This quantity physically represents the rate per unit area at which work is done by an element onto its neighbor. It can be also called `work flux' because it shows the work flow in the medium as well. When an element is compressed ($\bar{\sigma}>0$), it `pushes' its neighbor so that a positive work is done on the adjacent element. A notable fact is that there is a directionality to $I_2$, which depends on the direction of $v_1$.

Similarly, the cycle-average mechanical power $\langle I_2 \rangle$ can be expressed as $\langle I_2 \rangle=\tilde{I}e^{2\beta t}$, where 
\begin{align}
\tilde{I}=\frac{1}{2}\text{Re}[\hat{\bar{\sigma}}\hat{v}^*] \qquad\text{ [W/m$^2$]}.
\end{align}
The total mechanical power through the cross section $I$ of the rod is given by
\begin{align}
I=\int_A \langle I_2 \rangle dS=A\langle I_2 \rangle \qquad\text{ [W]}.
\end{align}

The work source can be further defined as the gradient of the mechanical power as
\begin{align}
w_2=\frac{\partial I_2}{\partial x} \qquad\text{ [W/m$^3$]}. \label{w2}
\end{align}
By expanding Eq. (\ref{w2}), $w_2$ can be further expressed as
\begin{align}
w_2=\frac{\partial\bar{\sigma}_1}{\partial x}v_1+\frac{\partial v_1}{\partial x}\bar{\sigma}_1
\end{align}
The first term of $w_2$ vanishes after applying cycle-averaging, because according to the momentum conservation (Eq. (\ref{1dmom})), $\partial \sigma_1/\partial x$ and $v_1$ are $90^\circ$ out of phase under the assumption that the small phase difference caused by the non-zero $\beta$ can be neglected due to: $\beta/\omega\ll 1$. The remaining term is equivalent to $\bar{\sigma}_1\frac{\partial \epsilon_1}{\partial t}$, \textit{i.e.}
\begin{align}
\frac{\partial v_1}{\partial x}\bar{\sigma}_1=\bar{\sigma}_1\frac{\partial \epsilon_1}{\partial t},
\end{align}
 whose cycle average is consistent with the cycle-averaged volume change work defined in \cite{Hao}.

The cross sectional integral of the work source is given by
\begin{align}
W=\int_A \langle w_2 \rangle dS=A\langle w_2 \rangle \qquad\text{ [W/m]}.
\end{align}

\begin{figure}[h]
	\centering
	\includegraphics[scale=0.4]{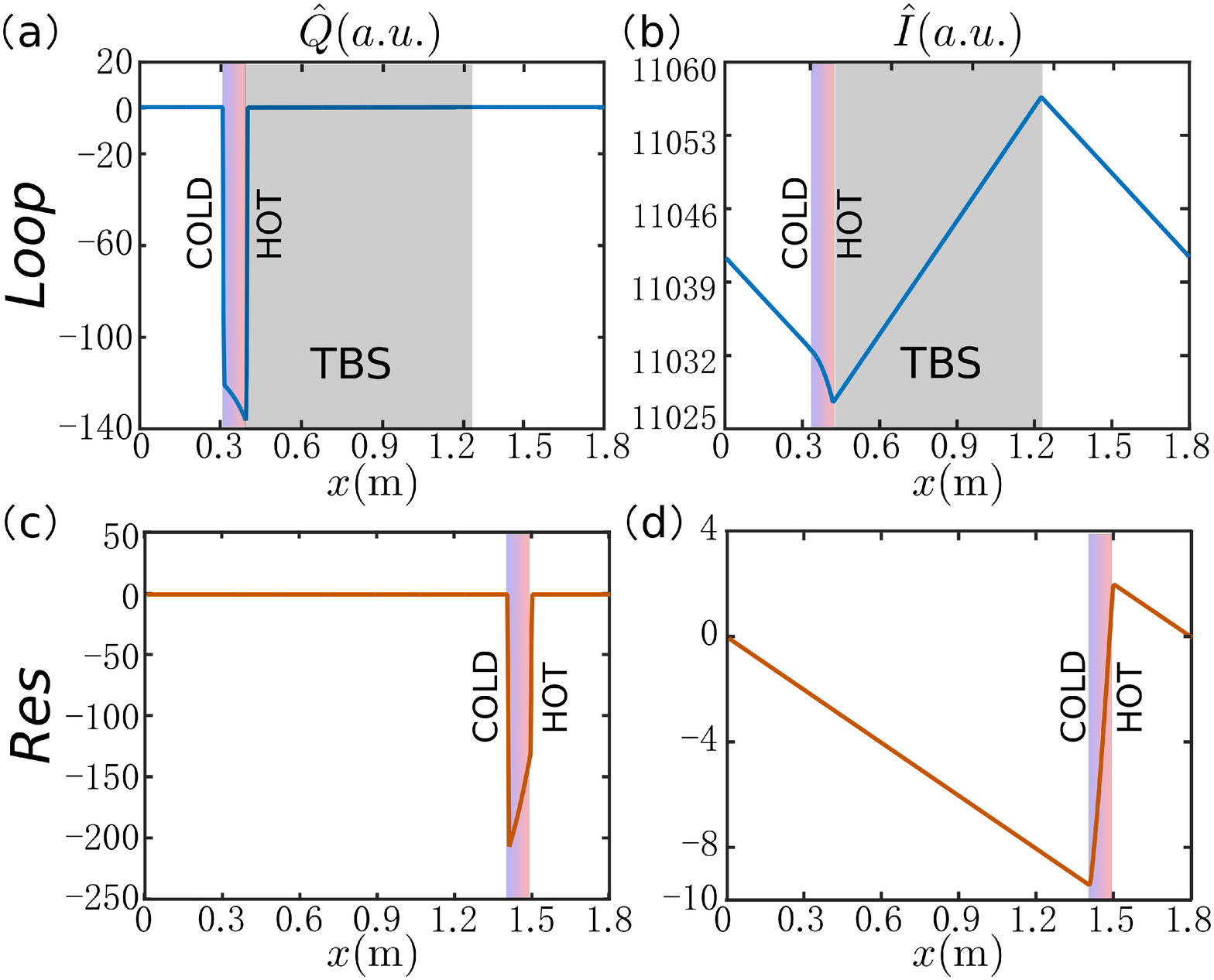}
	\caption{Cycle-averaged heat flux $\tilde{Q}$ and mechanical power $\tilde{I}$ in the frequency domain (arbitrary units) for the looped rod `$Loop$' and the resonance rod `$Res$', respectively. These components are evaluated from eigenfunctions from the stability analysis (Eqs.(\ref{1dint}), (\ref{1dmom}) and (\ref{1dene})). The color gradient strips indicate the location of S-segment, and the shaded grey strips indicate the location of the TBS in `$Loop$'. }
	\label{flux}
\end{figure}

Figure \ref{flux} shows the cycle-averaged quantities: heat flux $\tilde{Q}$ and mechanical power $\tilde{I}$ of a traveling wave engine (`$Loop$') and a standing wave one (`$Res$'). Note that the quantities indicated with $\tilde{()}$ satisfy the assumption of cycle averaging: $\langle ()_2 \rangle=\tilde{()}e^{2\beta t}$.  Figure \ref{flux}(a) and (c) illustrate that heat flux only exists in the S-segment and that wave-induced transport of heat occurs from the hot to the cold heat exchanger. The negative values in the S-segment in (a) and (c) are due to the fact that the hot exchanger is on the right side of the cold one, so heat flows to the negative $x$ direction in that case. The non-zero spatial gradient in $\tilde{Q}$ in the S-segment proves that there is heat exchange happening on the boundary of this segment because the heat flux in the axial direction is not balanced on its own. 

Fig. \ref{flux}(d) shows the mechanical power in the standing wave engine. The positive slope of $\tilde{I}$ in the S-segment elucidates the fact that the work generated in this region is positive, as discussed in detail in Section 5.2. This amount of work drops along the axial direction in the remaining segments at the spatial rate of $d\tilde{I}/dx$. The work drop in the hot and cold segments balances the accumulation of energy because there is no radial energy exchange in these sections. Clearly, if there is no energy growth, the slope of $\tilde{I}$ should be zero in these sections, as also discussed in Section 5.2.

The work flow in the traveling wave engine, as Fig. \ref{flux}(b) shows, has a very large value, which is due to the fact that negative stress $\bar{\sigma}$ and particle velocity $v$ have a phase difference much smaller than 90$^{\circ}$ (Fig. \ref{phase}). This means that a nearly uniform work flow is circulating the `$Loop$' carried by the wave dominated by TWC. Contrarily to the standing wave case, the slope of $\tilde{I}$ is negative in the S-segment, because it is balancing the positive work created by $\tilde{I}$ against the temperature gradient in the TBS. The volumetric integration of the work source $w$, \textit{i.e.} the spatial integration of $W$ along the rod, should be zero because, globally, their is no energy output in the system. All the energy converted from the heat in the S-segment should eventually lead to a uniformly distributed perturbation energy growth. More discussions will be addressed in the following section.

\subsection{\label{sec:level2}Acoustic Energy Budgets}
To derive the acoustic energy budgets, we first recast Eqs. (\ref{1dmom}) and (\ref{1dene}) in the time domain following the procedure by \cite{Gupta} as
\begin{align}
\frac{\partial v_1}{\partial t} &=-\frac{1}{\rho}\frac{\partial \bar{\sigma}_1}{\partial x} \label{E4},\\
\frac{\partial \bar{\sigma}_1}{\partial t} &=-E(1+\alpha \gamma_G T_0) \frac{\partial v_1}{\partial x}-\alpha E\frac{dT_0}{dx}v_1+\frac{\alpha E}{R\rho c_\varepsilon}q_1 \label{E5},
\end{align}
where, $q_1=2\kappa\frac{\partial T_1}{\partial r}\mid_{r=R}$ indicates the conductive heat flux at the medium-stage interface. 

Multiplying Eq. (\ref{E4}) by $\rho v_1$ and  Eq. (\ref{E5}) by $\bar{\sigma}_1 E^{-1}(1+\alpha \gamma_G T_0)^{-1}$, and adding them gives

\begin{align}
\frac{\partial \mathscr{E}_2}{\partial t}+ \frac{\partial I_2}{\partial x}+\mathscr{R}_2=\mathscr{P}_2-\mathscr{D}_2 \label{E6},
\end{align}
where
\begin{align}
\mathscr{E}_2&=\frac{1}{2}\rho v_1^2+\frac{1}{2}\frac{1}{E(1+\alpha \gamma_G T_0)}\bar{\sigma}_1^2, \label{E7}\\
I_2&=\bar{\sigma}_1v_1, \label{E8}\\
\mathscr{R}_2&=\frac{\alpha}{1+\alpha \gamma_G T_0}\frac{dT_0}{dx}I_2,\\
\mathscr{P}_2-\mathscr{D}_2&=\frac{\alpha}{1+\alpha \gamma_G T_0}\frac{1}{R\rho c_\varepsilon}q_1\bar{\sigma}_1. \label{E9}
\end{align}
$\mathscr{E}_2,I_2,\mathscr{R}_2, \mathscr{P}_2$ and $\mathscr{D}_2$ are the second order energy norm, work flux, energy redistribution term, thermoacoustic production and dissipation, respectively. Note that the work flux shown in Eq. (\ref{E8}) is consistent with the heuristic definition adopted in the previous section (Eq. (\ref{MP})).
With the harmonic convention $()_1=e^{(\beta+i\omega)t} \hat{()}$ and the assumption $\beta/\omega\ll1$, taking the cycle averaging of Eq. (\ref{E6}) yields
\begin{align}
2\beta{\tilde{\mathscr{E}}}+ \frac{d \tilde{I}}{d x}+\tilde{\mathscr{R}}=\tilde{\mathscr{P}}-\tilde{\mathscr{D}}, \label{E10}
\end{align}
where $\tilde{\mathscr{E}},\tilde{\mathscr{R}},\tilde{I},\tilde{\mathscr{P}}$, and $\tilde{\mathscr{D}}$ are transformed from the cycle averages of the cross-sectionally-averaged second order terms in Eqs. (\ref{E7}-\ref{E9}), following the assumption of cycle averaging: $\langle ()_2 \rangle=\tilde{()}e^{2\beta t}$. They are expressed as

\begin{align}
\tilde{\mathscr{E}}&=\frac{1}{2}\rho \vert\hat{v}\vert^2+\frac{1}{2}\frac{1}{E(1+\alpha \gamma_G T_0)}\vert\hat{\bar{\sigma}}\vert^2 quad&\text{ [W/m$^3$]},\\
\tilde{I}&=\frac{1}{2}\text{Re}[\hat{\bar{\sigma}}\hat{v}^*]\quad&\text{ [W/m$^2$]},\\
\tilde{\mathscr{R}}&=\frac{1}{2}\frac{\alpha}{1+\alpha \gamma_G T_0}\frac{dT_0}{dx}\text{Re}[\hat{\bar{\sigma}}\hat{v}^*] \quad&\text{ [W/m$^3$]},\\
\tilde{\mathscr{P}}&=\frac{1}{2}\frac{1}{1+\alpha \gamma_G T_0}\{\text{Re}[g_k]\text{Re}[\hat{\bar{\sigma}}(i\omega\hat{\varepsilon})^*+\text{Im}[g_k]\text{Im}[\hat{\bar{\sigma}}(i\omega\hat{\varepsilon})^*]\} \quad&\text{ [W/m$^3$]},\\
\tilde{\mathscr{D}}&=\frac{\omega}{2}\frac{1}{E(1+\alpha \gamma_G T_0)}\vert{\hat{\bar{\sigma}}}\vert^2 \text{Im}[g_k] \quad&\text{ [W/m$^3$]} \label{tildeD}.
\end{align}
The details of the derivations of Eqs. (\ref{E10})-(\ref{tildeD}) can be found in the supplementary material.

The growth rate can be recovered via
\begin{align}
\beta_\text{EB}=\frac{\tilde{\mathscr{P}}-\tilde{\mathscr{D}}-( \frac{\partial \tilde{I}}{\partial x}+\tilde{\mathscr{R}})}{2\tilde{\mathscr{E}}}. \label{E15}
\end{align}

\begin{figure}
	\centering
	\includegraphics[scale=0.8]{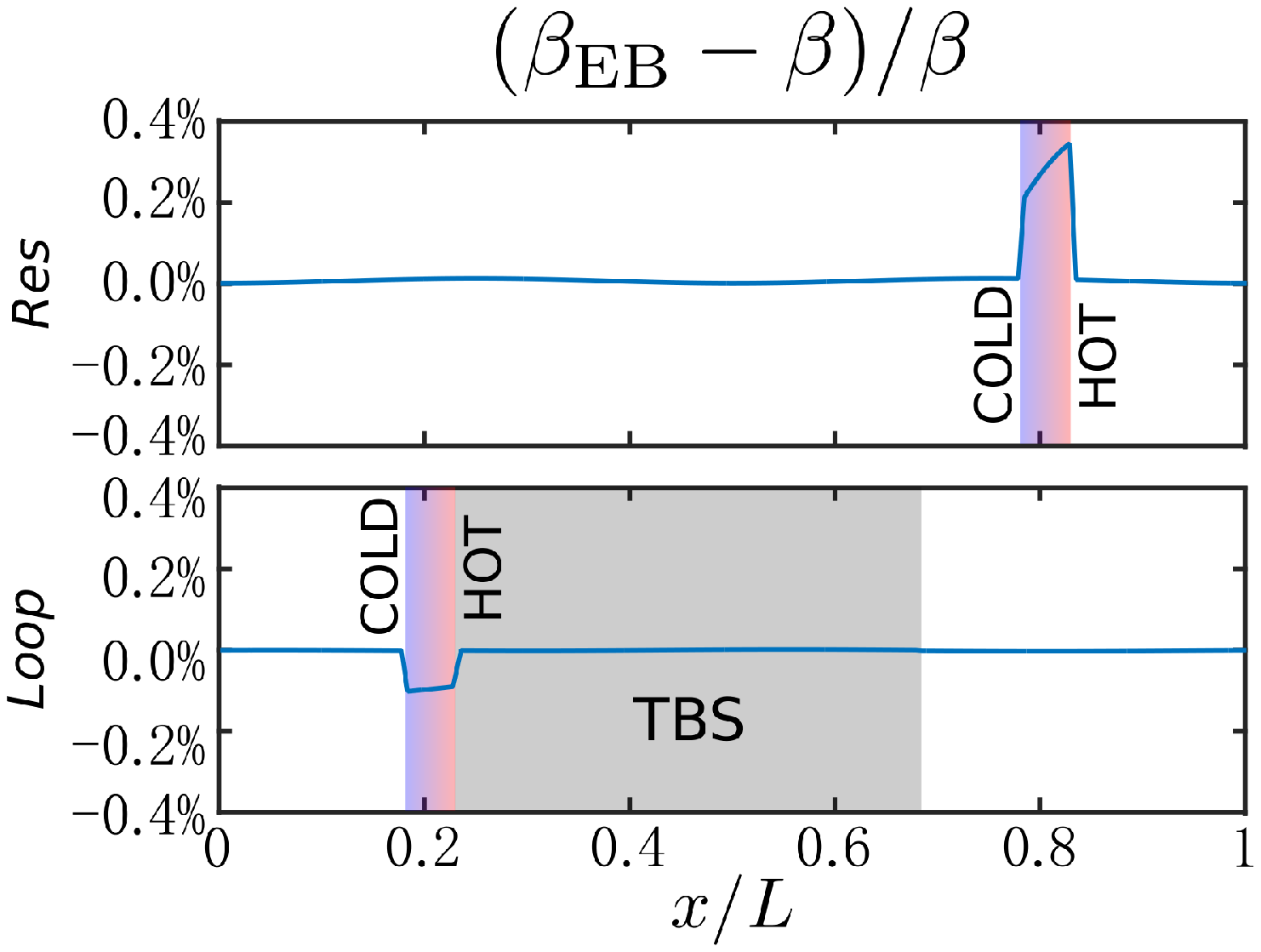}
	\caption{The relative difference of the growth rates estimated from the energy budgets $\beta_{\text{EB}}$ and directly retrieved from the eigenvalue problem in Eqs.  (\ref{1dint}), (\ref{1dmom}), and (\ref{1dene}) for the standing wave configuration (`$Res$') and the traveling wave configuration (`$Loop$').}
	\label{beta}
\end{figure}

As Fig. \ref{beta} shows, the growth rates $\beta_\text{EB}$ calculated from Eq. (\ref{E15}) are within $0.4\%$ from the direct output of the eigenvalue problem (Eqs. (\ref{1dint}), (\ref{1dmom}), and (\ref{1dene})) in both the standing wave and the traveling wave configurations, which validates the consistency of the derivations in this section. 

\begin{figure}
	\centering
	\includegraphics[scale=0.4]{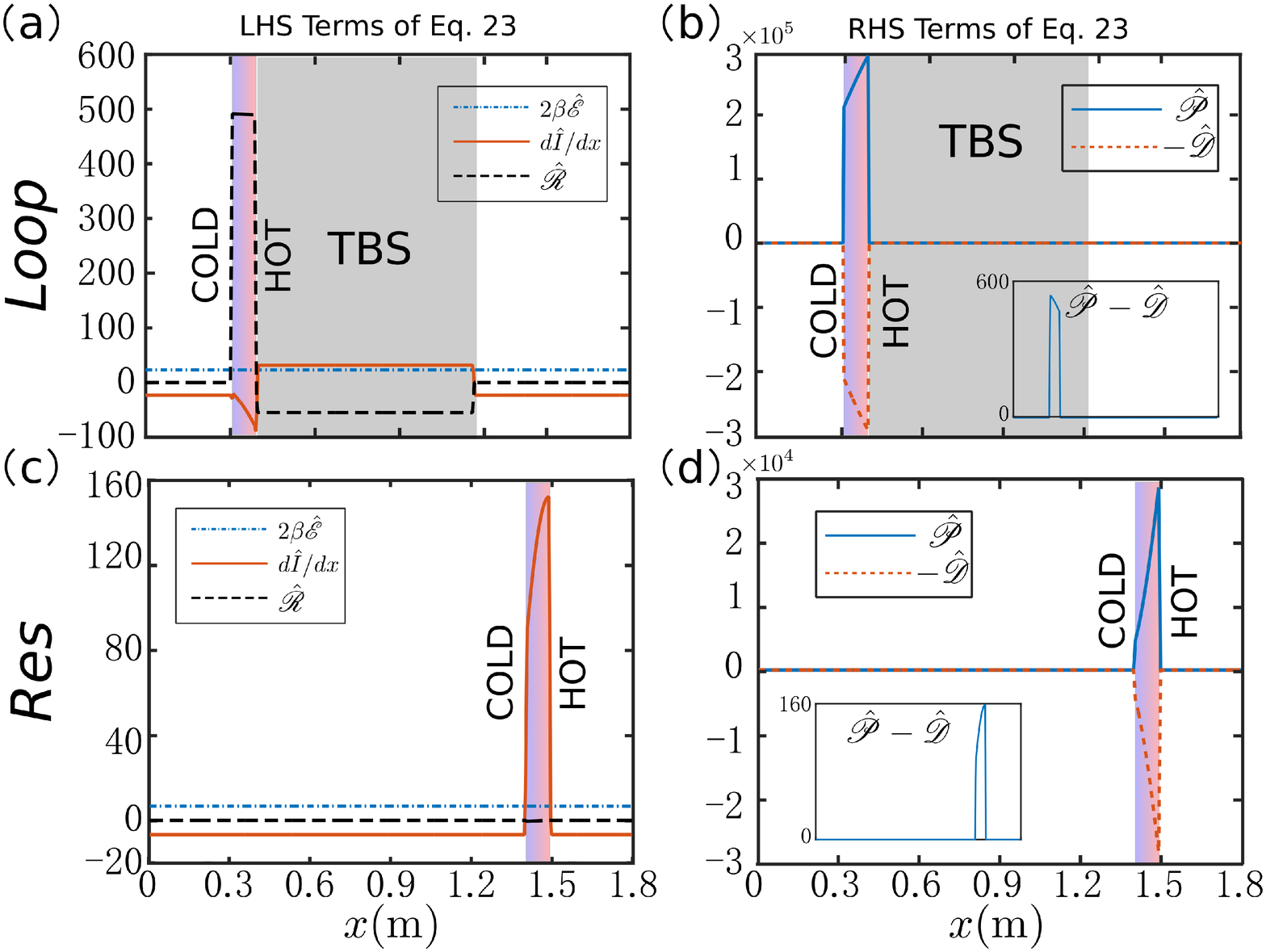}
	\caption{The terms in the acoustic energy budgets (Eq. (\ref{E10})) for (a) and (b) the traveling wave configuration (`$Loop$') and, (c) and (d) the standing wave configuration (`$Res$'). The insets in (b) and (d) plot the difference of the thermoacoustic production $\tilde{\mathscr{P}}$ and dissipation $\tilde{\mathscr{D}}$ in both configurations. The spatial integration of $\tilde{\mathscr{P}}-\tilde{\mathscr{D}}$ yields the total energy accumulation rate (see Eq. (\ref{intEB})).}
	\label{EB}
\end{figure}

From the physical point of view, the significance of the terms in Eq. (\ref{E10}) are illustrated as following. $2\beta{\tilde{\mathscr{E}}}$ quantifies the rate of energy accumulation, $d\tilde{I}/dx$ is the work source defined in the previous section, $\tilde{\mathscr{R}}$ is an energy redistribution term. $\tilde{\mathscr{P}}$ and $\tilde{\mathscr{D}}$ are the thermoacoustic production and dissipation, respectively. The energy redistribution term in the acoustic energy budgets of solid thermoacoustics cannot be found in the fluid counterpart of the same equations \cite{Gupta}. This term is absent in fluids because it is canceled in the algebraic derivations by expressing the variation of mean density according to the ideal gas law, as a function of the mean temperature gradient. On the other hand, in solid-state thermoacoustics, the heat-induced density variation is neglected and the impact of the temperature gradient is manifest in the stress-strain constitutive relation. It is proved numerically that the spatial integration of this term is zero (see Supplementary Material), so it does not produce or dissipate energy, but just \textit{redistributes} it. In summary, it represents the work created by the acoustic flux acting against the temperature gradient. Figure \ref{EB} plots every term in the acoustic energy budgets (Eq. (\ref{E10})) in the standing wave and traveling wave configurations, respectively.

The values of $\tilde{\mathscr{P}}$ and $\tilde{\mathscr{D}}$ are non-zero only in the S-segment. The dissipation $\tilde{\mathscr{D}}$ is due to wall heat transfer, which is a conductive loss. Although they are very similar in the S-segment, there exists a small difference between them. Thus, from a thermal standpoint, as a given amount of heat is transported through this section, a small portion of it (proportional to $\tilde{\mathscr{P}}-\tilde{\mathscr{D}}$) is converted into wave energy which accumulates in the rod, hence sustaining growth.

As can be seen, $2\beta\tilde{\mathscr{E}}$ is flat, meaning that the rate of the energy accumulation along the rod is uniform and exponential in time, consistent with the eigenvalue \textit{ansatz}. 

In the standing wave configuration, the work flux gradient $d\tilde{I}/d x$ peaks in the S-segment, and has a constant negative value out of the S-segment. As foreshadowed by the discussions in the previous section, this distribution means that $d\tilde{I}/d x$ adjusts itself so that $\beta$ is uniform. In other words, energy is accumulated everywhere at the same rate.

Neglecting the small phase shift caused by $\beta$, the energy redistribution  $\tilde{\mathscr{R}}$ does not exist in the standing wave configuration because of the $90^\circ$ phase difference between $\hat{\bar{\sigma}}$ and $\hat{v}$. Locally, the produced work in the S-segment, is converted from the most of the net production $\tilde{\mathscr{P}}-\tilde{\mathscr{D}}$. The remaining of $\tilde{\mathscr{P}}-\tilde{\mathscr{D}}$ transforms to the accumulated energy in this small segment. Outside the S-segment, the negative value of $d\tilde{I}/dx$ is exactly the same as the rate of the energy accumulation to keep the condition of zero local net production.

In the traveling wave configuration, the energy conversion becomes different because of the existence of the TBS. The TBS creates a temperature drop, which makes the energy redistribution term non zero in this section. To balance the negative value in the TBS, it peaks up in the S-segment so that the spatial integration is zero. In the TBS, the shape of the work flux gradient is the mirror image of that of the energy redistribution term because the addition of these two terms should be the negative of the spatially uniform energy accumulation rate. For the work flux gradient itself, a negative distribution in the S-segment is necessary to balance the positive redistributed work in the TBS so that the spatial integration is zero. The above supplements the explanations in the previous section on why the work source is negative in the S-segment.

Globally, in both configurations, given that both the spatial integrations of the work flux gradient and the energy redistribution terms are zero, the total net production $\int_0^L ({\tilde{\mathscr{P}}}-{\tilde{\mathscr{D}}}) dx$ only leads to the accumulation of energy
\begin{align}
\int_0^L 2\beta{\tilde{\mathscr{E}}} dx=\int_0^L ({\tilde{\mathscr{P}}}-{\tilde{\mathscr{D}}}) dx. \label{intEB}
\end{align}

\subsection{\label{sec:level2}Efficiency}
Generally, efficiency is defined as the ratio of work done to thermal energy consumed. However, since there is no energy harvesting element in the system, the rod has no work output. Thus, we take the accumulated energy, which could be potentially converted to energy output, as the numerator of the ratio. For the denominator, limited to the 1D assumption, the thermal energy consumed is not available directly from the quasi-1D model because the evaluation of the radial heat conduction at the boundary is lacking. Swift \cite{Swift2} suggested that the heat flux $\dot{Q}$ could be considered as uniform for a short stack, which is approximately equal to the consumed thermal energy. Thus, we use the averaged $\dot{Q}$ over the S-segment, an estimate of the consumed thermal energy, as the denominator of the efficiency. As a result, the efficiency $\eta$ is expressed as 

\begin{align}
\eta&=\frac{A\int_0^L \frac{\partial \mathscr{E}_2}{\partial t} dx}{\frac{1}{l_s} \int_{x_s-\frac{l_s}{2}}^{x_s+\frac{l_s}{2}}{\dot{Q}}dx}\\
&=\frac{\int_0^L 2\beta{\tilde{\mathscr{E}}} dx}{\frac{1}{l_s} \int_{x_s-\frac{l_s}{2}}^{x_s+\frac{l_s}{2}}\tilde{Q}dx}.
\end{align}

Although this definition is the best estimate we could make based on the quasi-1D model, we highlight that fully nonlinear 3D simulations are capable of providing more accurate estimates of the efficiency.

\begin{figure}
	\centering
	\includegraphics[scale=0.6]{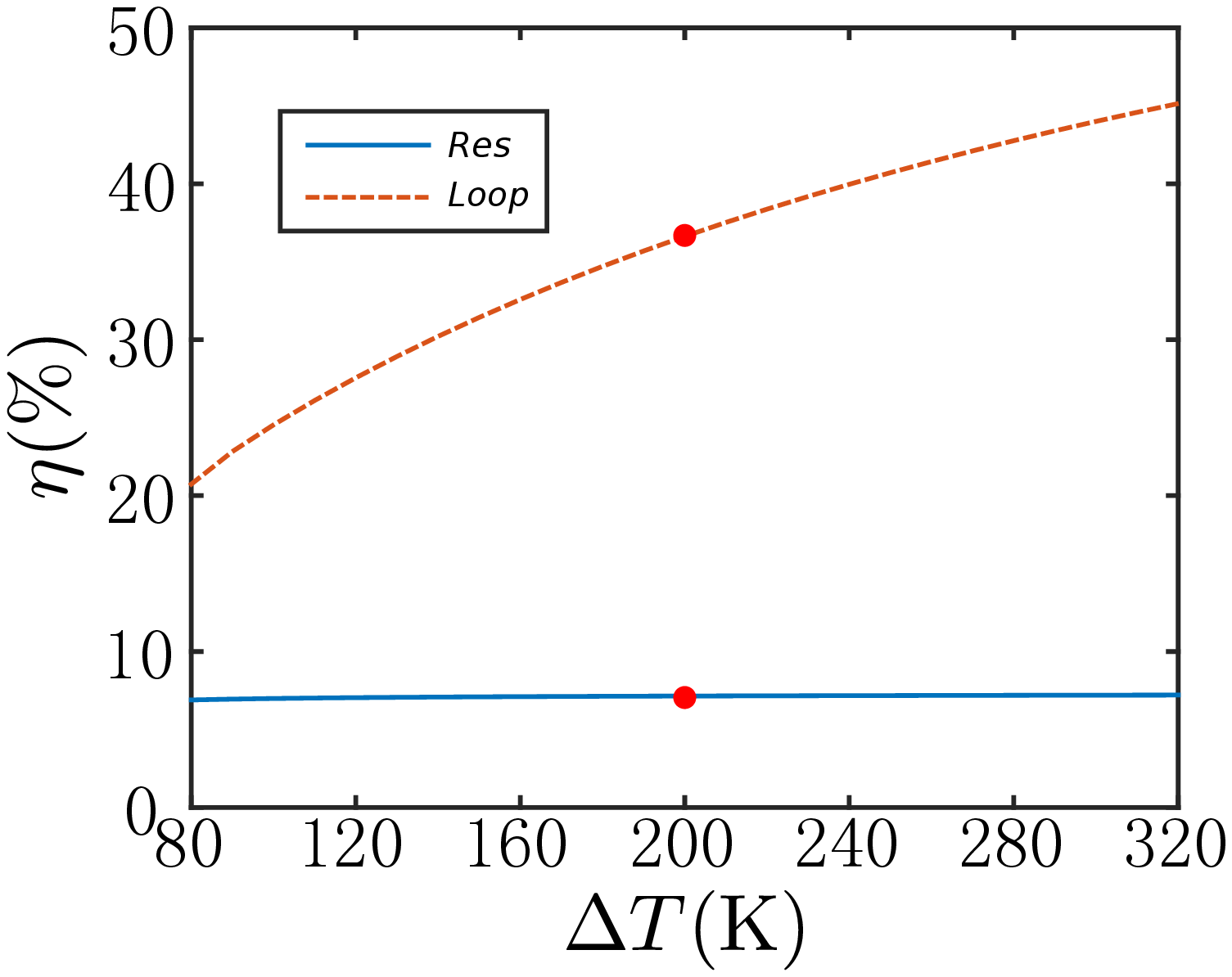}
	\caption{The efficiencies of the traveling wave configuration (`$Loop$') and the standing wave configuration (`$Res$') at different temperature difference $\Delta T$. The efficiencies are $37\%$ and $7\%$, respectively at $\Delta T=200K$ (The red dots). }
	\label{eta}
\end{figure}

Figure \ref{eta} shows the efficiencies of `$Loop$' and `$Res$' at different temperature difference $\Delta T=T_h-Tc$. It can be seen from this plot that (1) the efficiency of the traveling wave configuration `$Loop$' is much higher than that of the standing wave configuration $Res$, which is consistent with the conclusions drawn in fluids, and (2) for the traveling wave configuration, the efficiency goes up with $\Delta T$ increasing, while for the standing wave one, the efficiency is insensitive to the change of $\Delta T$. For the cases studied in the previous sections ($\Delta T=493.15\text{K}-293.15\text{K}=200\text{K}$), the efficiencies $\eta$ are $37\%$ and $7\%$ for `$Loop$' and `$Res$', respectively, as the red dots show in Fig. \ref{eta}.  

Considering that the material properties of solids are much more tailorable than fluids, the authors expect that the efficiency of SSTA can be improved by designing an inhomogeneous medium having optimized mechanical and thermal thermoacoustic properties.

\section{\label{sec:level1} Conclusions}
In this study, we have shown numerical evidence of the existence of traveling wave thermoacoustic oscillations in a looped solid rod. The growth ratio of a full wavelength traveling wave in a looped rod is found to be significantly larger than that of a full wavelength standing wave in a resonance rod. The phase delay in the looped rod between negative stress and particle velocity, which controls the value of TWC, is at most $30^\circ$ under the situation that the stage is $5\%L$ long and $\Delta T_0=200K$. Heat flux, mechanical power and work source are derived in analogous ways to their counterparts in fluids. The perturbation acoustic energy budgets are performed to interpret the energy conversion process of SSTA engines. The efficiency of SSTA engines is defined based on the rigorously derived energy budgets. The traveling wave SSTA engine is found to be more efficient than its standing wave counterpart. To conclude, this study confirms the theoretical existence of traveling wave thermoacoustics in a solid looped rod which could open the way to the next generation of highly-robust and ultra-compact traveling wave thermoacoustic engines and refrigerators.

\section{\label{sec:level1} Acknowledgments}
H. Hao would like to thank Prateek Gupta for the fruitful discussions and his helpful comments on acoustic energy budgets.

{\section{\label{sec:level1} Reference}
\bibliography{TWTAbib}

\begin{thebibliography}{10}
\expandafter\ifx\csname url\endcsname\relax
  \def\url#1{\texttt{#1}}\fi
\expandafter\ifx\csname urlprefix\endcsname\relax\def\urlprefix{URL }\fi
\expandafter\ifx\csname href\endcsname\relax
  \def\href#1#2{#2} \def\path#1{#1}\fi

\bibitem{Rayleigh}
Rayleigh, The explanation of certain acoustical phenomena, Nature 18 (1878) 319
  -- 321.

\bibitem{Poinsot}
T.~Poinsot, D.~Veynante, Theoretical and Numerical Combustion, R. T. Edwards,
  Inc., 2011.

\bibitem{Rijke}
P.~Rijke, {LXXI. N}otice of a new method of causing a vibration of the air
  contained in a tube open at both ends, Philos. Mag. Ser. 17(116) (1859) 419
  -- 422.

\bibitem{Hao}
H.~Hao, C.~Scalo, M.~Sen, F.~Semperlotti,
  \href{http://dx.doi.org/10.1063/1.5006489}{Thermoacoustics of solids: a
  pathway to solid state engines and refrigerators}, Journal of Applied Physics
  123~(2) (2018) 024903.
\newline\urlprefix\url{http://dx.doi.org/10.1063/1.5006489}

\bibitem{Swift2}
G.~Swift, \href{http://dx.doi.org/10.1121/1.396617}{Thermoacoustic engines},
  The Journal of the Acoustical Society of America 84~(4) (1998) 1145 -- 1180.
\newline\urlprefix\url{http://dx.doi.org/10.1121/1.396617}

\bibitem{Yazaki}
T.~Yazaki, A.~Iwata, T.~Maekawa, A.~Tominaga,
  \href{http://dx.doi.org/10.1103/PhysRevLett.81.3128}{Traveling wave
  thermoacoustic engine in a looped tube}, Physical Review Letters 81~(15)
  (1998) 3128 -- 31.
\newline\urlprefix\url{http://dx.doi.org/10.1103/PhysRevLett.81.3128}

\bibitem{Ceperley}
P.~Ceperley, \href{http://dx.doi.org/10.1121/1.383505}{A pistonless stirling
  engine-the traveling wave heat engine}, Journal of the Acoustical Society of
  America 66~(5) (1979) 1508 -- 13.
\newline\urlprefix\url{http://dx.doi.org/10.1121/1.383505}

\bibitem{Backhaus}
S.~Backhaus, G.~Swift, \href{http://dx.doi.org/10.1121/1.429343}{A
  thermoacoustic-stirling heat engine: Detailed study}, Journal of the
  Acoustical Society of America 107~(6) (2000) 3148 -- 66.
\newline\urlprefix\url{http://dx.doi.org/10.1121/1.429343}

\bibitem{Gedeon}
D.~Gedeon, \href{https://doi.org/10.1007/978-1-4615-5869-9_45}{DC Gas Flows in
  Stirling and Pulse Tube Cryocoolers}, Springer US, Boston, MA, 1997, pp.
  385--392.
\newline\urlprefix\url{https://doi.org/10.1007/978-1-4615-5869-9_45}

\bibitem{Ju}
Y.~L. Ju, C.~Wang, Y.~Zhou,
  \href{https://doi.org/10.1007/978-1-4757-9047-4_256}{Dynamic Experimental
  Study of the Multi-Bypass Pulse Tube Refrigerator with Two-Bypass Tubes},
  Springer US, Boston, MA, 1998, pp. 2031--2037.
\newline\urlprefix\url{https://doi.org/10.1007/978-1-4757-9047-4_256}

\bibitem{Ravex}
A.~Ravex, J.~M. Poncet, I.~Charles, P.~Bleuz{\'e},
  \href{https://doi.org/10.1007/978-1-4757-9047-4_247}{Development of Low
  Frequency Pulse Tube Refrigerators}, Springer US, Boston, MA, 1998, pp.
  1957--1964.
\newline\urlprefix\url{https://doi.org/10.1007/978-1-4757-9047-4_247}

\bibitem{Olson}
J.~Olson, G.~Swift,
  \href{http://dx.doi.org/10.1016/S0011-2275(97)00037-4}{Acoustic streaming in
  pulse tube refrigerators: tapered pulse tubes}, Cryogenics 37~(12) (1997) 769
  -- 776.
\newline\urlprefix\url{http://dx.doi.org/10.1016/S0011-2275(97)00037-4}

\bibitem{Boluriaan}
S.~Boluriaan, P.~Morris,
  \href{http://dx.doi.org/10.1260/147547203322986142}{Acoustic streaming: from
  rayleigh to today}, International Journal of Aeroacoustics 2~(3-4) (2003) 255
  -- 92.
\newline\urlprefix\url{http://dx.doi.org/10.1260/147547203322986142}

\bibitem{Scalo}
C.~Scalo, S.~Lele, L.~Hesselink,
  \href{http://dx.doi.org/10.1017/jfm.2014.745}{Linear and nonlinear modelling
  of a theoretical travelling-wave thermoacoustic heat engine}, Journal of
  Fluid Mechanics 766 (2015) 368 -- 404.
\newline\urlprefix\url{http://dx.doi.org/10.1017/jfm.2014.745}

\bibitem{Swift}
G.~Swift, \href{http://dx.doi.org/10.1121/1.403896}{Analysis and performance of
  a large thermoacoustic engine}, Journal of the Acoustical Society of America
  92~(3) (1992) 1551 -- 63.
\newline\urlprefix\url{http://dx.doi.org/10.1121/1.403896}

\bibitem{Rott}
N.~Rott, \href{http://dx.doi.org/10.1007/BF01595562}{Damped and thermally
  driven acoustic oscillations in wide and narrow tubes}, Zeitschrift fur
  Angewandte Mathematik und Physik 20~(2) (1969) 230 -- 43.
\newline\urlprefix\url{http://dx.doi.org/10.1007/BF01595562}

\bibitem{Yates}
B.~Yates, Thermal Expansion, Plenum Press, New York, 1972.

\bibitem{Biot}
M.~Biot, \href{http://dx.doi.org/10.1063/1.1722351}{Thermoelasticity and
  irreversible thermodynamics}, Journal of Applied Physics 27~(3) (1956) 240 --
  253.
\newline\urlprefix\url{http://dx.doi.org/10.1063/1.1722351}

\bibitem{Gupta}
P.~Gupta, G.~Lodato, C.~Scalo,
  \href{http://dx.doi.org/10.1017/jfm.2017.635}{Spectral energy cascade in
  thermoacoustic shock waves}, Journal of Fluid Mechanics 831 (2017) 358 --
  393.
\newline\urlprefix\url{http://dx.doi.org/10.1017/jfm.2017.635}

\end{thebibliography}

\end{document}